\documentclass[prl, floats, aps, prd, preprint, tightenlines, groupedaddress, nofootinbib, showpacs, byrevtex, superscriptaddress, preprintnumbers]{revtex4}

\usepackage{amsfonts}
\usepackage{amssymb,latexsym}
\usepackage{amsmath,amsbsy}
\usepackage{epsfig,bm}
\usepackage{graphicx,comment}
\def\cO{{\mathcal O}}
\def\cV{{\mathcal V}}

\begin{document}

\preprint{JLAB-THY-05-411}
\vphantom{}

\title{PQChPT with Staggered Sea and Valence
Ginsparg-Wilson Quarks: Vector Meson Masses\\[10pt]}

\author{Hovhannes R.~Grigoryan}
 \email{hgrigo1@jlab.org}
 \affiliation{Thomas Jefferson National Accelerator Facility, 12000
              Jefferson Ave., Newport News, VA 23606, USA\\}

 \affiliation{Louisiana State University, Department of Physics \&
              Astronomy, 202 Nicholson Hall, Tower Dr., LA 70803, USA}

 \affiliation{Laboratory of Theoretical Physics, JINR, Dubna, Russian
              Federation\\
 \qquad}

\author{Anthony W.~Thomas}
 \email{awthomas@jlab.org}
 \affiliation{Thomas Jefferson National Accelerator Facility, 12000
              Jefferson Ave., Newport News, VA 23606, USA\\}

\vskip 0.2cm
\begin{abstract}
\vskip 0.2cm
We consider partially quenched, mixed chiral perturbation theory
with staggered sea and Ginsparg-Wilson valence quarks in order to
extract a chiral-continuum extrapolation expression for the vector
meson mass up to order $ \cO(a^2) $, at one-loop level. Based on
general principles, we  accomplish the task without explicitly
constructing a sophisticated, heavy vector meson chiral Lagrangian.
\end{abstract}



\volumeyear{year}
\volumenumber{number}
\issuenumber{number}
\eid{identifier}

\maketitle

\section{Introduction}

In the past few years there has been significant progress in the
study of the chiral-continuum extrapolation problem for lattice QCD
with different types of fermions (see, for example,
Refs.~\cite{Bar:2002nr, Bar:2003mh, Bar:2005tu, Tiburzi:2005is}).
Extrapolation of lattice QCD simulation results to the physical
(light) quark masses is a nontrivial task due to the nonanalytic
variation of hadronic properties with quark masses which is a
consequence of spontaneous chiral symmetry breaking. To perform an
extrapolation to the continuum limit we need to account for direct
chiral, taste (in the staggered case) and rotational symmetry
breaking by finite lattice spacing.

At present, lattice QCD simulations with staggered fermions defined
in Ref.~\cite{Susskind} reached smaller quark masses
\cite{Bernard:2001av} compared with other types of lattice fermions.
For example, compared with unquenched simulations using
Ginsparg-Wilson fermions (defined in Ref.~\cite{Ginsparg})
simulations with staggered fermions are computationally less
demanding. The computational cost depends on many factors like
lattice spacing and the practical implementation of the fermions in
the simulations (see Ref.~\cite{Bar:2005tu} and references therein).
From Ref.~\cite{Kennedy:2004ae} it follows that the simulations with
Ginsparg-Wilson fermions may be about ten to one hundred times more
expensive than with improved staggered fermions at comparable
masses.

However, the price of these advantages of staggered fermions is that
they contain an internal flaw. In those simulations, to reduce the
number of taste degrees of freedom, the fourth-root trick is
employed. In the continuum limit this trick is consistent because
there are no taste-changing interactions, however, when one includes
these interactions the finite lattice spacing makes the root of the
determinant non-local and the trick remains controversial (see, for
example, Ref.\cite{Bunk:2004br}). In order to include the relatively
large discretization effects in staggered calculations, staggered
chiral perturbation theory has been developed (see
Refs.\cite{Lee:1999zx, Aubin:2003mg, Aubin:2003rg, Sharpe:2004is}).
Application of the staggered Lagrangian allowed one to determine the
pseudoscalar masses and decay constant \cite{Aubin:2003mg,
Aubin:2003rg, VandeWater:2005uq}. For example, recently in
Ref.~\cite{Gray:2005ad} there has been considerable success in
determining the $ B $ meson decay constant using the MILC
collaboration unquenched gauge configurations with three flavors of
light sea quarks. In these calculations both valence and sea light
quarks were represented by the highly improved (AsqTad) staggered
quark action which allows for a much smoother chiral extrapolation
to physical up and down quarks than has been possible in the past.

Being computationally very costly, Ginsparg-Wilson fermions, on the
other hand, do not suffer from the inconsistency of staggered
fermions. Moreover, these fermions have an exact chiral symmetry. It
has been shown in Refs.~\cite{Bar:2005tu, Tiburzi:2005is} that it is
possible to combine the good chiral properties of Ginsparg-Wilson
fermions, taking them as valence quarks where they are
computationally less demanding, with numerically fast staggered
fermions as sea quarks. Extrapolation for the hybrid (mixed fermion)
program has recently been addressed for baryons by Tiburzi in
Ref.~\cite{Tiburzi:2005is}. Here we consider the application of this
program for the $ \rho $ meson, which is a classic testing ground for
QCD with light quarks.\\

In this report we extend the study of QCD for staggered sea and
valence Ginsparg-Wilson quarks in Ref.\cite{Bar:2005tu} to calculate
the mass of the vector meson at one-loop level up to order $
\cO(a^2) $. QCD with staggered sea and valence Ginsparg-Wilson
quarks, which as in \cite{Bar:2003mh, Bar:2002nr} we will refer to
as mixed QCD, consists of $ N_v $ fully quenched (valence) quarks
(with $ N_v $ ghost-quarks) and $ 4 N_f $ staggered (sea) quarks.
The masses of the sea and valence quarks, in principle, could be
chosen to be completely arbitrary, except that the quenched quark
masses are fixed to be equal to the masses of their ghost partners.
The full graded chiral symmetry of mixed QCD in the chiral and
continuum limits is described via the semi-direct product $ G_{MQCD}
= SU(4N_f + N_v|N_v)_L \times SU(4N_f + N_v|N_v)_R \times U(1)_V $,
where $ 4N_f $ and $ N_v $ are the numbers of unquenched (staggered)
and quenched quarks, respectively. As in full QCD, we expect that
the axial $ U(1)_A $ symmetry is also broken by the anomaly. We also
phenomenologically include in our calculations the one-loop
contribution from the $ \rho \rightarrow \pi\pi $ decay which cannot
be consistently incorporated in the heavy meson formalism because of
the high momentum involved in this process. However, the inclusion
of this loop is justified by the physical results obtained, for
example, in Refs.~\cite{Allton:2005fb, Grigoryan:2005zj,
Leinweber:2001ac} and references therein.

\section{Symanzik Action}

In order to consistently include lattice discretization effects the
Symanzik action \cite{Symanzik:1983dc, Symanzik:1983gh} needs to be
constructed, based on the symmetry constraints of the underlying
lattice theory, with further projection to the low energy effective
theory. This task was carried out in Refs.\cite{Bar:2005tu,
Tiburzi:2005is} and here, we will only briefly mention the main
results.

In general, the Symanzik action up to order $ \cO(\epsilon^4) $ has
the form:
\begin{align} \label{Symanzik}
S_{sym} = S_{PQ} + a S_1 + a^2 S_2 + a^3 S_3 + a^4 S_4 + \ldots
\end{align}
\noindent where $ S_{PQ} $ is the action corresponding to continuum
partially quenched theory described in Refs.~\cite{Morel,
Bernard:1993sv}, $ S_n $ ($ n \in N $) is the action consisting of
dimension $ (n + 4) $ operators. Here, we will require that the
heavy meson chiral Lagrangian, corresponding to $ S_{PQ} $, matches
the corresponding partially quenched chiral Lagrangian, similar to
that in Ref.~\cite{Chow:1997dw} or in Ref.~\cite{Grigoryan:2005zj}
when the lattice spacing is set to zero. From discussions in
Ref.~\cite{Tiburzi:2005is} (see also Refs.~\cite{Luscher:1998pq,
Sharpe:1993ng, Luo:1996vt}) the actions $ S_1 $ and $ S_3 $ are
equal to zero because of the specific character of the fermions we
use\footnote{There are no dimension 5(7) operators which can be
consistently formed by the quark bilinear and which preserve the
symmetries of the theory in the approximation we are interested in,
see Ref.~\cite{Tiburzi:2005is} and references therein.}. The action
$ S_2 $, according to Ref.~\cite{Bar:2005tu}, contains only six
allowed operators which break taste or rotational symmetries (see
also for details Refs.~\cite{Bar:2003mh,Tiburzi:2005is,SW}). The
chiral Lagrangian corresponding to the action $ S_4 $ will give only
a tree level contribution of order $ \cO(a^4) $. Although allowed in
our counting scheme, this contribution will be disregarded, as
currently we are interested in the $ \cO(a^2) $ analytic behavior of
vector meson masses.

Using the results from the operator construction in
Refs.~\cite{Bar:2005tu, Tiburzi:2005is} we will study the analytic
behavior of the vector meson mass in both the chiral and continuum
limits, without constructing the explicit expression for the heavy
meson PQChPT.

\section{Goldstone Meson Multiplet Sector}

Mixed ChPT in the Goldstone meson sector with staggered and
Ginsparg-Wilson quarks was first studied in Ref.\cite{Bar:2005tu}.
We first give a brief description of that work.\\[-10pt]

Assuming that the $ G_{MQCD} $ symmetry is spontaneously broken down
to its vector part, the particle spectrum will contain light
pseudoscalar bosons. These Goldstone meson fields can be written in
terms of a $ (4N_f+2N_v)\times(4N_f+2N_v)$, unitary matrix field, $
\Sigma $, defined as:
\begin{align}
\Sigma = exp(2\imath\Phi/f),
\end{align}
where $ \Sigma \in U(4N_f+N_v|N_v) $, $ \Phi $ is the bosonic matrix
and $ f $ is a low energy constant. In particular, we are interested
in the case with $ N_f = 3 $ and $ N_v = 2 $, for which the bosonic
matrix has the following form:
\begin{equation} \label{Phi}
\Phi = \left(
         \begin{array}{ccccccc}
           U        & \pi^+ & K^+            & Q_{ux} & Q_{uy} & ... & ...
           \\[5pt]
           \pi^-    & D     & K^0            & Q_{dx} & Q_{dy} & ... & ... \\[5pt]
           K^-      & \bar{K}^0 & S        & Q_{sx} & Q_{sy} & ... & ... \\[5pt]
           Q^\dag_{ux} & Q^\dag_{dx}    & Q^\dag_{sx} & X & P^+ & R^\dag_{\tilde{x}x} & R^\dag_{\tilde{y}x} \\[5pt]
           Q^\dag_{uy} & Q^\dag_{dy}    & Q^\dag_{sy} & P^- & Y & R^\dag_{\tilde{x}y} & R^\dag_{\tilde{y}y} \\[5pt]
           ...      & ...         & ...      & R_{\tilde{x}x}   & R_{\tilde{x}y}   & \tilde{X} & \tilde{P}^+ \\[5pt]
           ...      & ...         & ...      & R^\dag_{\tilde{y}x} & R^\dag_{\tilde{y}y} & \tilde{P}^- & \tilde{Y} \\[5pt]
         \end{array}
       \right)
\end{equation}
\noindent Following the notation of Ref.~\cite{Bar:2005tu}, the
following correspondence is implied: $ P^+ \rightarrow x\bar{y} $, \
\ $ X \rightarrow x\bar{x} $, \ \ $ Y \rightarrow y\bar{y} $, \ \ ($
\tilde{P}^+ $, $ \tilde{X} $ and $ \tilde{Y} $ are the analogous
combinations of valence ghost quarks), $ R_{\tilde{x}x} \rightarrow
\tilde{x}\bar{x} $, similarly for $ R_{\tilde{x}y} $, $
R_{\tilde{y}x} $, and $ R_{\tilde{y}y} $. $ Q_{Fv} \rightarrow
F\bar{v} $, where $ F \in \{u, d, s\} $ and $ v \in \{x, y\} $ (note
that $ Q_{Fv} $ is a $ 4 \times 1 $ matrix in taste). The
pseudoscalar states $ U $, $ D $, $\pi^+$,..., etc. in
Eq.~(\ref{Phi}), which consist of staggered quarks only, form the
``staggered'' sector of the mixed pseudoscalar boson ChPT. This
sector is well studied in Refs.~\cite{Aubin:2003mg, Aubin:2003rg,
Sharpe:2004is}. Each state in the ``staggered'' sector can be
represented as:
\begin{align}
U = \sum^{16}_{a = 1} U_a \frac{T_a}{2}, \ \ \  T_a = \{\xi_5,
\imath \xi_{\mu 5}, \imath\xi_{\mu \nu}, \xi_{\mu}, \xi_{I} \}
\end{align}
\noindent (similarly for $ D $, $ K^+ $, etc ) where $ T_a $ are the
sixteen Euclidean gamma matrices associated with taste (see, for
example, Ref.\cite{Aubin:2003mg}).

\noindent Under chiral symmetry transformations:
\begin{align}
\Sigma  \rightarrow L \Sigma R^\dag,
\end{align}
where $ L(R) \in SU(4N_f + N_v |N_v )_{L(R)} $. As in
Refs.~\cite{Bar:2005tu, Tiburzi:2005is}, we adopt the following
counting scheme:
\begin{align}
m_q/\Lambda_{QCD} \approx a^2 \Lambda^2_{QCD} \cong \epsilon^2,
\end{align}
where $ \Lambda_{QCD} $ denotes the typical QCD scale. This counting
scheme is relevant for simulations with improved staggered quarks
\cite{Aubin:2004fs}. The leading order chiral Lagrangian (in this
counting scheme) that contains the terms of order $ \cO(p^2,m_q,
a^2) $ is of the form:
\begin{equation}\label{Lagrangian}
L_{\Phi} = \frac{f^2}{8} <\partial_{\mu}\Sigma
\partial^{\mu}\Sigma^\dag> - \ \frac{f^2B}{4}<\Sigma M^\dag  + \ \Sigma^\dag
M > + \ \frac{m^2_0}{6} <\Phi^2> + \ a^2 \cV.
\end{equation}
\noindent Here $ <...> $ denotes a supertrace in flavor space and $
B $ is a low-energy constant. In our case the diagonal mass matrix
is:
\begin{equation}
 M = diag(m_u \xi_I, m_d \xi_I, m_s \xi_I, m_x, m_y, m_x, m_y).
\end{equation}
For convenience, to derive the flavor neutral propagators, the
singlet mass parameter $ m^2_0 $ in Eq.~(\ref{Lagrangian}) is
allowed to be finite, however, at the end of the calculations we
always could take the limit  $ m^2_0 \rightarrow \infty $.

The potential $ \cV $ in the leading order Lagrangian comprises all
terms proportional to $ a^2 $. At this stage we will not go deep
into describing the form of the $ \cV $ (we refer to
Refs.~\cite{Bar:2005tu, Aubin:2003mg} for more information). Here,
we present only those results required for our further study.

At tree level, because of the specific properties of potential $ \cV
$, the valence-valence mesons with only Ginsparg-Wilson quarks obey
the continuum-like mass relations:
\begin{align}\label{val-val}
 m^2_P = B(m_x + m_y), \    \   \   \ m^2_{X(Y)} = 2B m_{x(y)}.
\end{align}
In our calculations the only relevant valence-valence Goldstone
bosons will be flavor neutral because the valence quarks in the sea
loops will be canceled by their ghost partners and, in addition,
G-parity suppresses the process\footnote{$ \rho $ and $ \eta' $ have
even G-parity, while the pseudoscalar meson has odd G-parity} $ \rho
\rightarrow \eta'\pi $. We also will be interested only in
flavor-neutral and taste singlet staggered quark combinations, so in
the sea-sea sector, for a meson in a singlet taste channel $ I $
made of sea quarks of flavor $ F $, we have:
\begin{equation}\label{sea-sea}
m^2_{F_I} = 2Bm_F + a^2 \Delta_I
\end{equation}
\noindent where the constant $ \Delta_I $ is given, for example, in
Ref.~\cite{Bar:2005tu} (Eq.~(27)). Finally, the masses of the
valence-sea mesons ($ F\bar{v} $ or $ \bar{F}v $) are given as:
\begin{align}\label{mixed}
m^2_{Q_{Fx(y)}} = B(m_F + m_{x(y)}) + a^2 \Delta_{Mix},
\end{align}
\noindent where $ \Delta_{Mix} $ is an undetermined parameter in
this mixed theory which is taste independent.

The flavor-charged (non-diagonal) fields in Eq.~(\ref{Phi}) have
only connected propagators ($ G^{con}_{ij} $), in the quark-flow
sense, which are given by the expression\footnote{we will work only
with Euclidean propagators}:
\begin{align}
G^{con}_{ij}(p) = \frac{\delta_{ij} \epsilon_i}{p^2 + m^2_{ii}}
\end{align}
\noindent where $ i $, $ j $ $ \in $ \{ghost, quark\}, $ \epsilon_i
= 1 $ if $ i = $ quark and $ - 1 $ if $ i =$ ghost. Disconnected
propagators for flavor neutral mesons (like, for example, $ U $ or $
X $) contain hairpin-like interactions (through the $ m^2_0 $ term).
The sum of all these hairpin-like diagrams (in the way described in
Ref.\cite{Aubin:2003mg}) for the disconnected $ XY $ propagators ($
G^{disc}[XY] $) will give (as in Ref.~\cite{Bar:2005tu}) the
following result:
\begin{align}
G^{disc}[XY] = -\frac{m^2_0}{3}\frac{(q^2 + m^2_{U_I})(q^2 +
m^2_{D_I})(q^2 + m^2_{S_I})}{(q^2 + m^2_{X})(q^2 + m^2_{Y})(q^2 +
m^2_{\pi^0_I})(q^2 + m^2_{\eta_I})(q^2 + m^2_{\eta'_I})}
\end{align}
\noindent where $ \pi^0_I $ , $ \eta_I $ and $ \eta'_I $ are the
diagonalized flavor-neutral, singlet taste-channel mass eigenstates.
Although there should be vector and axial-vector hairpin
contributions, as in Ref.~\cite{Aubin:2003mg}, we do not include
them here because they are irrelevant in our later calculations.

\section{Mixed One-Loop Corrections to the Vector Meson Mass}

ChPT with staggered sea and chiral (Ginsparg-Wilson) valence quarks
was successfully applied to calculate the corrections to
pseudoscalar masses and decay constants to one loop in
Ref.~\cite{Bar:2005tu} and recently, in Ref.~\cite{Tiburzi:2005is},
this approach was applied to calculate masses and magnetic moments
of baryons. In this work we will calculate vector meson mass
corrections at one-loop level, without explicitly constructing the
corresponding heavy meson Lagrangian but rather using general principles.\\[-7pt]

From Eq.~(\ref{Symanzik}) it follows that continuum heavy meson
chiral Lagrangian will consist of the parts: $ L_1 \sim \cO(p) \sim
\cO(\epsilon) $ and $ L_2 \sim \cO(m_q) \sim \cO(\epsilon^2) $. As
shown in Ref.~\cite{Grigoryan:2005zj}, from the comparison with the
simulations in Ref.~\cite{Allton:2005fb}, the contribution from the
terms of order $ \cO(m^2_q) $ may be neglected\footnote{Although
Ref.~\cite{Grigoryan:2005zj} considers effective theory with
different types of fermions, the results in the continuum limit
should match.}. The chiral Lagrangian, $ L_b $, corresponding to $
S_2 $ in Eq.~(\ref{Symanzik}), will consist of terms of order $
\cO(a^2) \sim \cO(\epsilon^2) $. And finally, we will not have terms
of order $ \cO(p^{n + 1}) $ ($ n \in N $) in the heavy meson chiral
Lagrangian because they can be eliminated by using the equations of
motion \cite{Arzt:1993gz}.

Now, it is easy to see that $ L_1 $ gives one-loop chiral
corrections to the vector meson mass -- see the diagrams $ (a,c,d) $
shown in Fig.~1. The terms in $ L_2 $ and $ L_b $ will give tadpole
as well as tree level contributions. From these considerations for
the total tree level correction $ \sigma_{tree} $ to the vector
meson mass (up to order $ \cO(\epsilon^4) $) it follows that:
\begin{align}
\sigma_{tree}(\mu) = c_1 a^2 + c_2 m_x + c_3 m_y + c_4 m_u + c_5 m_d
+ c_6 m_s
\end{align}
\noindent where, in general, $ c_i = c_i(\mu) $ ($ i = 1,...,6 $)
with $ \mu $ as the renormalization scale.

For concreteness we will calculate the mass correction of a charged
$ \rho $ meson at rest. (The calculation is somewhat more
complicated for a neutral $ \rho $ meson, but because of isospin
symmetry, the mass correction of all $ \rho $ mesons must be the
same.)

\begin{figure}[!ht]
\begin{center}
\includegraphics[scale = 0.3]{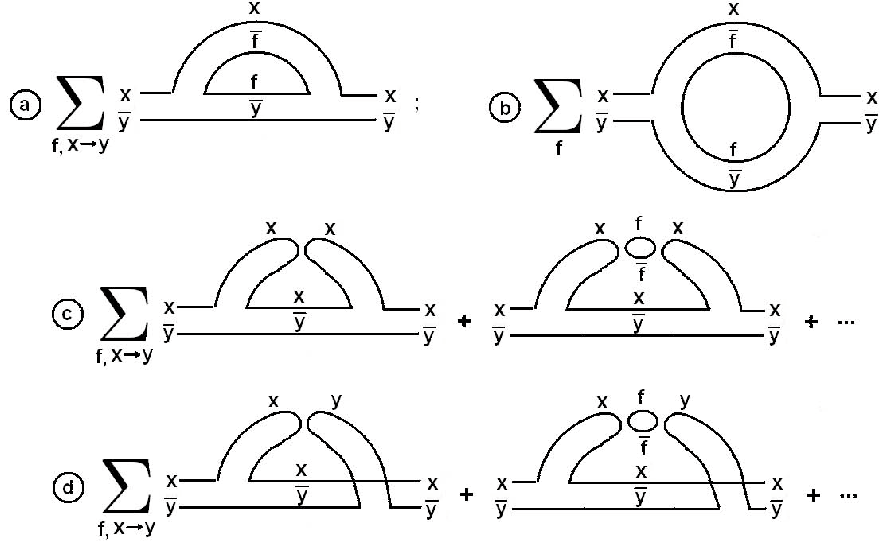}
\caption{The quark flow diagrams}
\end{center}
\end{figure}

In the case $ N_f = 3 $ and $ N_v = 2 $ we have staggered $ u $, $ d
$ and $ s $ sea quarks with $ x $ and $ y $ valence quenched quarks
(corresponding to different flavors). The quark flow diagrams for
this case are presented in Fig.~1, with the propagators for the
connected diagrams (Fig.~1 $a,b$) given by:
\begin{align}
\nonumber G^{con}_{vvp} &= \sum_{f = u,d,s} \left[\frac{1}{(q^2 +
m^2_{Q_{fx}})} +
\frac{1}{(q^2 + m^2_{Q_{fy}})}\right] \frac{1}{((p-q)^2 + M^2)} \ , \\
G^{con}_{vpp} &= \sum_{f = u,d,s} \frac{1}{(q^2 +
m^2_{Q_{fx}})((p-q)^2 + m^2_{Q_{fy}})},
 \end{align}
\noindent where $ p $ is the initial momentum of the vector meson
with mass $ M $ ($ p^2 = - M^2 $ in Euclidean space), $ q $ is the
momentum of intermediate pseudoscalar meson state and instead of
using a heavy meson propagator we just used the usual one (which is
practically the same) with the mass of intermediate vector meson
equal to the mass of the initial one. The subindices on the
propagators remind us that in the self-energy integral we need to
take into account the vector-vector-pseudoscalar ($ vvp $) and
vector-pseudoscalar-pseudoscalar ($ vpp $) nature of the meson
interactions. For the diagrams in Fig.~1$c$ and $d$ in the limit $
m_0 \rightarrow \infty $, we have:
\begin{align}
G^{DHP}_{vvp} = -\frac{1}{3}\frac{(q^2 + m^2_{U_I})(q^2 +
m^2_{D_I})(q^2 + m^2_{S_I})}{(q^2 + m^2_{\pi^0_I})(q^2 +
m^2_{\eta_I})((p-q)^2 + M^2)}\left[\frac{1}{(q^2 + m^2_{X})} +
\frac{1}{(q^2 + m^2_{Y})}\right]^2.
 \end{align}
\noindent Notice, that for case shown in Fig.~1$b$ there are no
additional double hairpin diagrams because of G-parity. Now, let us
take $ m_u = m_d \equiv m $ ($ N_f = 2 + 1 $ picture) as in MILC
simulations \cite{Aubin:2004wf}. In this case:
\begin{align}
\nonumber m^2_{U_I} = m^2_{D_I} = m^2_{\pi^0_I}\\
m^2_{\eta_I} = \frac{1}{3} m^2_{U_I} + \frac{2}{3}m^2_{S_I}
 \end{align}
\noindent  Let us also require that $ m_x = m_y $ and $ m_{U_I} =
m_X $ or equivalently $ B m_x = B m + \frac{1}{2} a^2 \Delta_{I} $.
The first choice will help us to find the correction to the $ \rho $
meson mass because it consists of two light quarks. The second
choice is natural because in the continuum limit it gives $ m_x = m
$. Taking these two conditions into account we can write the above
propagator in the form:
\begin{align}
G^{DHP}_{vvp} = - \frac{2}{3}\left[ \frac{3}{q^2 + m^2_{X}} -
\frac{1}{q^2 + m^2_{\eta_I}} \right] \frac{1}{((p-q)^2 + M^2)}.
 \end{align}
\noindent Using the method in Ref.~\cite{Pichowsky:1999mu} for the
total self-energy we have:
 \begin{align}
   \Sigma_{total} &= \sum_{f = u,d,s}\left( 2\Sigma_{vvp}(M^2,m^2_{Q_{fx}}, \Lambda) +
   \Sigma_{vpp}(m^2_{Q_{fx}}, \Lambda)\right) \\ \nonumber &- 2\Sigma_{vvp}(M^2,m^2_{X}, \Lambda) + \frac{2}{3}
   \Sigma_{vvp}(M^2,m^2_{\eta_I}, \Lambda)
 \end{align}
\noindent where
    \begin{align}
   \Sigma_{vvp}(M^2, m^2, \Lambda) &=
   -\frac{g^2M}{12\pi^2}\int^\infty_0 dk\frac{k^4u^2_{vvp}(k^2, \Lambda) }{\omega^2},\\[10pt]
   \Sigma_{vpp}(m^2, \Lambda) &= -\frac{f^2_{\rho \pi \pi}}{6\pi^2}\int^\infty_0 dk \frac{k^4 u^2_{vpp}(k^2, \Lambda)}{\omega(\omega^2 -
   \frac{M^2}{4})},
    \end{align}
\noindent we used the notation $ \omega^2 = k^2 + m^2 $ and the
functions $ u^2_{vvp(vpp)}(k^2, \Lambda) $ are finite range
regulators (\cite{Pichowsky:1999mu, Leinweber:2001ac,
Donoghue:1998bs,Detmold:2001hq}). For example, the leading
non-analytic (LNA) contribution to the self energy is:
    \begin{align}
   \Sigma^{LNA}_{total} &= -\frac{g^2M}{24\pi}\left(4m^3_{Q_{ux}} + 2m^3_{Q_{sx}}
   - 2m^3_{X} + \frac{2}{3}m^3_{\eta_I} \right)
    \end{align}
\noindent In the continuum limit (where we have $ m_x = m $) the LNA
contribution is:
 \begin{align}
   \delta M^{LNA} &= \frac{\Sigma^{LNA}_{total}}{2M} = -\frac{g^2}{48\pi}
   \left(2m^3_{\pi} + 2 m^3_{K} + \frac{2}{3}m^3_{\eta}\right).
 \end{align}
\noindent This result agrees with that in Ref.~\cite{Jenkins:1995vb}
for the vector $ \rho $ meson - where in the limit $ N_c \rightarrow
\infty $ they found:
 \begin{align}
  \delta M^{LNA}_{\rho} = - \frac{g^2_2}{12 \pi f^2} \left[ 2 m^3_{\pi} + 2 m^3_{K} + \frac{2}{3}m^3_{\eta}
  \right].
 \end{align}
\noindent Note that while the coefficients in these two expressions
are written differently, since $ g = 2g_2/f $ for values of $ g
\equiv g_{\rho \omega \pi} = 16$ $GeV^{-1}$
(\cite{Leinweber:2001ac}), $ g_2 = 0.75 $ (\cite{Jenkins:1995vb})
and $ f = 93.75 $ $MeV$, the expressions are equal.

It would be also interesting to consider the case, when $ m_u = m_d
$, $ m_X = m_{U_I} $ and $ m_Y = m_{S_I} $ ($ m_y = m_s $ in the
continuum limit). Using these conditions and recalling that the $
K^*$ vector meson consists of one light and one strange quark we
find the self energy contribution:
 \begin{align}\label{sunrise}
   \Sigma_{total} = \sum_{f = u,d,s}\left( \Sigma_{vvp}(M^2,m^2_{Q_{fx}},\Lambda)
   + \Sigma_{vvp}(M^2,m^2_{Q_{fy}},\Lambda) + \Sigma_{vpp}(m^2_{Q_{fx}},m^2_{Q_{fy}},\Lambda)\right)
   - \\ \nonumber
   - \frac{1}{2}\Sigma_{vvp}(M^2,m^2_{X},\Lambda) - \Sigma_{vvp}(M^2,m^2_{Y},\Lambda) + \frac{1}{6} \Sigma_{vvp}(M^2,m^2_{\eta_I},\Lambda)
 \end{align}
\noindent In the continuum limit ($ m_x = m_u $ and $ m_y = m_s $)
the LNA contribution to the total self-energy is:
 \begin{align}
   \Sigma^{LNA}_{total} = \frac{g^2_2}{12 \pi f^2} \left(\frac{3}{2}m^3_{\pi} + 3m^3_{K} +
   \frac{1}{6}m^3_{\eta}\right).
 \end{align}
\noindent Again, comparing this result with that in
Ref.~\cite{Jenkins:1995vb} for the $ K^* $ meson mass correction (in
the limit $ N_c \rightarrow \infty $):
 \begin{align}
  \delta M_{K^*} = - \frac{g^2_2}{12 \pi f^2} \left[ \frac{3}{2}m^3_{\pi} + 3 m^3_{K} + \frac{1}{6}m^3_{\eta} \right]
 \end{align}
\noindent we see good agreement with our result for the LNA
contribution in the continuum limit.

Now, returning to $ \rho $ meson, let us take into account the
tadpole contribution from $ L_{2(b)} \sim \cO(a^2, m_q) $. This case
is similar to $ L_1 $ but without intermediate vector meson
propagators and we don't need to include the diagram with two
intermediate pseudoscalars. The total self-energy correction from $
L_{2(b)} $ is:
 \begin{align}\label{tadpole}
   \widetilde{\Sigma}_{total} = 4\Sigma_{tad}(m^2_{Q_{ux}},\Lambda) + 2\Sigma_{tad}(m^2_{Q_{sx}},\Lambda)
   - 2\Sigma_{tad}(m^2_{X},\Lambda) +
   \frac{2}{3}\Sigma_{tad}(m^2_{\eta_I},\Lambda),
 \end{align}
\noindent with the definition:
 \begin{align}
    \Sigma_{tad}(m^2,\Lambda) = \tilde{\beta}\int \frac{d^4 k}{(2\pi)^4}
    \frac{k^2 u^2(k^2, \Lambda)}{(k_0 - \imath\epsilon)^2(k^2 + m^2 +
    \imath\epsilon)},
 \end{align}
\noindent where, in general, $ \tilde{\beta} = \beta' m^2 + \beta
a^2 $, $ \beta' $ and $ \beta $ are undetermined parameters and $
u(k^2, \Lambda) $ is a finite range regulator
\cite{Leinweber:2001ac,Donoghue:1998bs,Detmold:2001hq}. The term
proportional to $ \beta' $ is absent in the continuum limit (see,
for example, Ref.~\cite{Grigoryan:2005zj} when lattice spacing is
set to zero). This means that the tadpole contribution comes only
from the operators in $ L_b $. Then, for example, if $ u = 1 $
(applying dimensional regularization) we have:
 \begin{align}
  \Sigma_{tad}(m^2) = \frac{\beta a^2}{16\pi^2}m^2 ln
  \frac{m^2}{\mu^2} \sim \cO(\epsilon^4).
 \end{align}

The calculations above were performed in the case where $ x $ and $
y $ correspond to different flavors. To calculate the mass of the
neutral vector meson we need to change our quark flow diagrams and
add one more with DHP insertions on each intermediate pseudoscalar
line. However, as noted above, isospin symmetry for the vector
meson, means that we need not perform separate calculations for the
neutral vector meson mass.

\section{Vector Meson Masses up to order $ \cO(\epsilon^4) $}

Imposing the conditions $ m_x = m_y = m_u + \frac{1}{2B}a^2
\Delta_{I} $ with $ m_u = m_d $ we can simplify the expression for
the tree level contribution and the vector meson mass in both chiral
and continuum expansions can be written in the form:
\begin{align}
M^2_V = \left(c + c_a a^2 + c_q m^2_{\pi} + c_s m^2_{S_I}\right)^2 +
\Sigma_{total}(\mu) + \widetilde{\Sigma}_{total}(\mu),
\end{align}
\noindent where $ m^2_{\pi} = 2B m_x = m^2_X $, $ m^2_{S_I} = 2B m_s
+ a^2 \Delta_I $ and $ c $, $ c_{a,q,s} $ are new dimensionful
parameters. The dependence on $ m^2_{S_I} $ may be absorbed in the
coefficients $ c $ and $ c_a $ if one fixes the mass of the strange
quark in the simulations.

From the definition of $ \Sigma_{vvp} $ and $ \Sigma_{vpp} $ using
the appendix in Ref.~\cite{Leinweber:2001ac} we can separate the
leading non-analytic contributions from $ \Sigma_{total} $ and $
\widetilde{\Sigma}_{total} $
\begin{align}
   \nonumber \Sigma^{LNA}_{total} &= -\frac{\mu_{\rho}g^2}{12\pi}\left( 2m^3_{Q_{ux}} + m^3_{Q_{sx}} - m^3_{X} +
   \frac{1}{3}m^3_{\eta_I} \right) \\  \ \ \ & \ \ \ \ \ \ \ \ \ \ \ \ \ \ - \frac{f^2_{\rho \pi \pi}}{4\pi^2 \mu^2_{\rho}} \left[2m^4_{Q_{ux}}ln \frac{m_{Q_{ux}}}{\mu} +
   m^4_{Q_{sx}} ln \frac{m_{Q_{sx}}}{\mu} \right], \\[20pt]
   \nonumber \widetilde{\Sigma}^{LNA}_{total} &= \frac{\beta a^2}{8\pi^2}\left[2m^2_{Q_{ux}}ln \frac{m_{Q_{ux}}}{\mu} +  m^2_{Q_{sx}}ln \frac{m_{Q_{sx}}}{\mu}
   - m^2_{X}ln \frac{m_{X}}{\mu} + \frac{1}{3}m^2_{\eta_I}ln \frac{m_{\eta_I}}{\mu}
   \right]
 \end{align}
\noindent where $ \mu_{\rho} $ is the physical mass of the $ \rho $
meson and using Eqs.~(\ref{val-val}--\ref{mixed}) we can express all
meson masses as functions of $ m^2_{\pi} $ and $ m^2_{S_I} $:
\begin{align}
m^2_{Q_{ux}} &= m^2_{\pi} + a^2 \left(\Delta_{Mix} + \frac{1}{2}\Delta_{I}\right),\\
m^2_{Q_{sx}} &=  \frac{1}{2}m^2_{\pi} + \frac{1}{2}m^2_{S_I} + a^2 \left(\Delta_{Mix} -
\frac{1}{2}\Delta_{I}\right),\\[5pt]
m^2_{\eta_I} &= \frac{1}{3}m^2_{\pi} + \frac{2}{3}m^2_{S_I}.
\end{align}
It is also of practical interest to consider the general case with
only two conditions: $ m_u = m_d $ and $ m_x = m_y $. In this case
the total self-energies are:
 \begin{align}
   \nonumber \Sigma_{total} &= \sum_{f = u,d,s}\left(2\Sigma_{vvp}(M^2,m^2_{Q_{fx}},\Lambda)
   + \Sigma_{vpp}(m^2_{Q_{fx}},\Lambda)\right)
   \\ \nonumber
   &- \frac{4}{3}\alpha\Sigma_{vvp}(M^2,m^2_{X},\Lambda) -
   \frac{4}{3}(1 -\alpha)\Sigma_{vvp}(M^2,m^2_{\eta_I},\Lambda) +
   \frac{4}{3}\tilde{\alpha}\frac{\partial\Sigma_{vvp}(M^2,m^2_{X},\Lambda)}{\partial
   m^2},
 \end{align}
 \begin{align}
   \nonumber \widetilde{\Sigma}_{total} &= 4\Sigma_{tad}(m^2_{Q_{ux}},\Lambda) + 2\Sigma_{tad}(m^2_{Q_{sx}},\Lambda)
   \\ \nonumber
   & - \frac{4}{3}\alpha\Sigma_{tad}(m^2_{X},\Lambda) -
   \frac{4}{3}(1 -\alpha)\Sigma_{tad}(m^2_{\eta_I},\Lambda) +
   \frac{4}{3}\tilde{\alpha}\frac{\partial\Sigma_{tad}(m^2_{X},\Lambda)}{\partial
   m^2},
 \end{align}
where
 \begin{align}
   \nonumber \alpha &= 1 - \frac{m^2_{\eta_I} - m^2_{S_I}}{m^2_{\eta_I} - m^2_{X}} +
   \frac{(m^2_{\eta_I} - m^2_{S_I})(m^2_{U_I} - m^2_{X})}{(m^2_{\eta_I} - m^2_{X})^2} \\[10pt]
   \nonumber \tilde{\alpha} &= (m^2_{U_I} - m^2_{X})\left[1 - \frac{m^2_{\eta_I} - m^2_{S_I}}{m^2_{\eta_I} - m^2_{X}}\right]
 \end{align}
from which one can see if $ m_X = m_{\pi} = m_{U_I} $ then we have
the same result as in Eqs.~(\ref{sunrise},\ref{tadpole}).

To perform chiral-continuum extrapolation of lattice data for the $
\rho $ meson with Ginsparg-Wilson quarks in a staggered sea the
consistent numerical evaluation of self-energy integrals $
\Sigma_{total}(\mu) $ and $ \tilde{\Sigma}_{total}(\mu) $ needs to
be done using the particular type of finite range regulator. The
extrapolation formula, in a numerical simulation with fixed strange
quark mass, will have five coefficients, some of which are functions
of the FRR cutoff.

\section{Conclusion}

Based on the recently developed mixed PQChPT for pseudoscalar meson
sector we have extracted a chiral-continuum extrapolation formula
for the vector meson masses at one-loop up to order $ \cO(a^2) $.
Instead of writing explicit expression for the vector meson
Lagrangian we have used general principles in order to calculate the
mass corrections at one-loop order.

Up to order $ \cO(\epsilon^4) $, vector meson masses, in general,
depend on six, presently unknown, parameters: $ c $, $ c_{a,q,s} $,
$ \beta $ and $ \Delta_{Mix} $. The low energy coefficient $
\Delta_{Mix} $ is model independent and may also be estimated from
the extrapolation of pseudoscalar masses. If one, however, is
interested in vector meson mass corrections up to order $
\cO(\epsilon^3) $ then the number of parameters reduces by two:
\begin{align}
M^2_V = \left(c + c_a a^2 + c_q m^2_{\pi} + c_s m^2_{S_I}\right)^2 -
\frac{\mu_{\rho}g^2}{12\pi}\left( 2m^3_{Q_{ux}} + m^3_{Q_{sx}} -
m^3_{X} + \frac{1}{3}m^3_{\eta_I} \right).
\end{align}
This can be used to test the theory predictions up to order $
\cO(\epsilon^3) $. It is also possible to reduce the number of
extrapolation parameters by one if we fix the strange quark mass in
the simulations.

During the derivations the following approximations were used: $ m_x
= m_y $, $ m_u = m_d $ and $ m_{X} = m_{U_I} $ for the $ \rho $
meson. This approximation may be adjusted, for example, to the $ K^*
$ vector meson if we assume $ m_u = m_d $, $ m_{X} = m_{U_I} $ and $
m_{Y} = m_{S_I} $. The idea beyond this is that in the continuum
limit the masses of the valence quarks, $ m_x $ and $ m_y $, are
required to correspond to the masses of valence quarks from which
the vector meson is constructed.

In our calculations we include (phenomenologically) the relativistic
$ \rho \rightarrow \pi\pi $ decay process, which although cannot be
consistently included in the framework of low energy effective field
theory, is required to give consistent physical predictions.
Finally, we did not take into account the finite volume effects from
the lattice, which are already well studied in
Refs.~\cite{Arndt:2004bg,Bedaque:2004dt,Orth:2005kq,Colangelo:2004sc}.\\[20pt]

\section*{Acknowledgements}

H.R.G. would like to thank Jerry P. Draayer for his support, the
Southeastern Universities Research Association (SURA) for a Graduate
Fellowship and Louisiana State University for a Scholarship
partially supporting his research. He would also like to acknowledge
S. Vinitsky and G. Pogosyan for support at the Joint Institute of
Nuclear Research. {}Finally, this work was supported in part by DOE
contract DE-AC05-84ER40150, under which SURA operates Jefferson
Laboratory.

\end{document}